\documentclass[12pt]{article}
\usepackage{amsmath,mathrsfs}
\usepackage{float,dsfont,multirow}
\usepackage{amssymb,mathtools,amsthm,bm,bbm}
\usepackage{enumitem,enumerate}
\usepackage{hyperref,natbib,url}
\usepackage{verbatim,epsfig,amsbsy,color}
\usepackage{subfigure,graphicx,geometry,tabularx}
\usepackage{appendix,algorithm,algorithmic}
\usepackage{cancel,soul,ulem}
\usepackage{nameref,zref-xr}
\usepackage[modulo]{lineno}

\newcommand{\cov}{\mathrm{Cov}}
\newcommand{\E}{\mathbb{E}}
\newcommand{\const}{\mathrm{c}}

\theoremstyle{remark}

\newtheorem{theorem}{Theorem}

\newtheorem{proposition}{Proposition}

\newtheorem{assumption}{Assumption}

\newcommand{\blind}{1}

\begin{document}
\setulcolor{red}
\setstcolor{red}

\def\spacingset#1{\renewcommand{\baselinestretch}%
{#1}\small\normalsize} \spacingset{1}


\if1\blind
{
  \title{\bf Covariance test for discretely observed functional data: when and how it works?}
  \author{Yang Zhou \\
  School of Statistics,
  Beijing Normal University, Beijing, China \\
  Jin Yang\\
  Department of Applied Mathematics\\
  The Hong Kong Polytechnic University, Hong Kong\\
  Fang Yao \thanks{
    Fang Yao is the corresponding author: fyao@math.pku.edu.cn. Yang Zhou and Jin Yang are co-first authors, and contributed equally.
}\\
    Department of Probability \& Statistics, School of Mathematical Sciences,\\
    Center for Statistical Science, Peking University, Beijing, China}
    \date{}
  \maketitle
} \fi

\if0\blind
{
  \bigskip
  \bigskip
  \bigskip
  \begin{center}
    {\LARGE\bf Covariance Test for Discretely Observed Functional Data: When and How It Works?}
\end{center}
  \medskip
} \fi

\begin{abstract}
For covariance test in functional data analysis, existing methods are developed only for fully observed curves, whereas in practice, trajectories are typically observed discretely and with noise.
To bridge this gap, we employ a pool-smoothing strategy to construct an FPC-based test statistic, allowing the number of estimated eigenfunctions to grow with the sample size.
This yields a consistently nonparametric test, while the challenge arises from the concurrence of diverging truncation and discretized observations.
Facilitated by advancing perturbation bounds of estimated eigenfunctions, we establish that the asymptotic null distribution remains valid across permissable truncation levels. Moreover, when the sampling frequency (i.e., the number of measurements per subject) reaches certain magnitude of sample size, the test behaves as if the functions were fully observed.
This phase transition phenomenon differs from the well-known result of the pooling mean/covariance estimation, reflecting the elevated difficulty in covariance test due to eigen-decomposition.
The numerical studies, including simulations and real data examples, yield favorable performance compared to existing methods.
\end{abstract}

\vfill

\newpage
\spacingset{1.9} 

\section{Introduction}
Functional data, referring to data generated from continuous underlying processes, are collected in many scientific fields such as medical imaging, speech recognition, molecule biology and environmental meteorology, etc. Functional data analysis (FDA), a popular branch of statistics over the past decades, has attracted considerable attention, see \cite{Ramsay05FDA} for an introduction and \cite{Hsing15FDA} for some theoretical foundations.
The research on hypothesis testing for functional data is mostly concentrated on mean functions \citep{Fan98test,Horvath13twomeanfts,Horvath15equalmeanfts} with scarce attention paid to covariance operators, which lay the foundation of subsequent analysis such as functional principal component (FPC) and regression analysis \citep[][among others]{Yao05fda,Yao05flr,Hall06eigen,Hall06dispca,Hall07flr}, and is the focus of this article. To motivate the proposed testing method, we point out that the challenges to be encountered are two-fold. Functional covariance test is intrinsically a nonparametric test on compact operators that reside in the infinite-dimensional space, and is substantively complicated by discretely observed and noise-contaminated data in practice.

For two-sample covariance test, a typical line of existing methods aims to metrizing the difference directly based on appropriate metric/distance (e.g., Hilbert-Schmidt norm), to construct test statistics whose asymptotic distributions depend on infinite-dimensional parameters which are unknown and difficult to estimate in practice. Consequently, these fully-functional tests suffer intensive calculations by some resampling methods (e.g., permutation and/or bootstrap) to obtain approximate critical values \citep[see][for example]{Pigoli14distancecov,Paparoditis16bootstrap,Boente18multiplecovtest}.
Another line of research exploited the projection-based methods, defining test statistics on a finite-dimensional space, where the basis is commonly formed by a fixed number of eigenfunctions \citep{Panaretos10eqaulcov,Kraus12,Horvath13equalcov}. Facilitated by dimension reduction, the derived asymptotic distributions of FPC-based tests are more tractable due to the finite-dimensional approximation, however, the fixed truncation adopted in these works violated the nonparametric nature.
It is noteworthy that all methods in literature have been only developed for fully observed functional data that may be viewed as an ``ideal'' scenario. Recently, \cite{Kraus19} investigated the $k$-sample test for covariance operators when functional data are partially observed and established the asymptotic distributions of FPC-based tests for fixed truncations.

The ``practical'' scenario is that one observes $n$ random functions at $N_i$ $(1\leq i\leq n)$ discrete time points with noise contamination, where $N_i$ may vary for each subject and is often assumed $N_i\asymp N$. Here $a_n\asymp b_n$ if $a_n\gtrsim b_n$ and $b_n\gtrsim a_n$, where $a_n\gtrsim b_n$ indicates $a_n\geq\const b_n$ for some constant $\const$ and large $n$. There are two typically smoothing strategies when dealing with such discretely observed functional data, one of which is only applicable to sufficiently dense data by pre-smoothing individual curves before further analysis. For sparse functional data, it is suggested to borrow information from all subjects by pooling observations together to estimate population parameters \citep{Yao05fda}. Here the magnitude of $N$, referred to as the sampling frequency,  plays an important role in the choice of smoothing methods and the asymptotic behavior with a phenomenon called ``phase transition''. For example, \cite{ZJT07stainf} and \cite{Kong16pflr} have shown that, when $N\gtrsim n^{5/4}$ the mean and covariance estimators based on pre-smoothed curves achieve the parametric $\sqrt{n}$-convergence. In contrast, the pool-smoothing estimators for the mean and covariance functions attain optimal $\sqrt{n}$-consistency when $N \gtrsim n^{1/4}$ \citep{Cai11dismeanRKHS,Zhang16sparse}. These results provide theoretical insights into the advantages of pool-smoothing estimation over the pre-smoothing method. Nevertheless, the above-mentioned phase transition are only valid for mean and covariance estimation, while for rigorous inference, especially involving eigenfunctions based on discretely observed data, there has not been meaningful progress to our knowledge. Viewing ubiquitousness of discretely observed data, we are inspired to bridge the gap from the discretized observations to the continuum and answer the key question: when and how a covariance test procedure works?

Despite the possibility of using pre-smoothing or pool-smoothing methods to conduct fully-functional tests, it is challenging to study the theoretical properties due to individual smoothing errors or the complex structure of pool-smoothing estimators, especially when considering a unified framework for functional data from sparse to dense designs. FPC-based tests could effectively capture differences that commonly arise from significant variations across the spectra of covariance operators due to good behaviour of the eigen-system, and their explicit asymptotic distributions could be derived, laying the theoretical foundation for conducting statistical inference. Thus we adopt FPC-based tests in this paper and the number of eigen-projections (i.e., the truncation level) is not fixed but rather allowed to increase with the sample size. This flexible approach places our method within a nonparametric regime, enabling us to enhance statistical power against deviations in any direction with increasing sample size.

To bridge the gap between the continuum and discretized observations, a crucial step is to quantify the perturbation bounds for a diverging number of estimated eigenfunctions based on discretely observed data, which is vital for deriving the asymptotic test distribution. A seminal work of \cite{Hall07flr} has established sharp bounds for estimated eigenfunctions with diverging index based on the cross-sectional sample covariance that is only applicable for fully observed curves. For discretely observed functions, ranging from sparse to dense designs, it is advantageous to consider the pooling smoothing strategy given its effectiveness in mean/covariance estimation as discussed earlier. However, the perturbation technique in \cite{Hall07flr} is no longer applicable, although the one-dimensional nonparametric rate has been obtained for a fixed number of estimated eigenfunctions based on pooling smoothing \citep{Hall06dispca}. Recently, \cite{ZWY22flrdis} has made meaningful progress on perturbation bounds for a diverging number of estimated eigenfunctions based on pooling smoothing of discretely observed data. This enables us to make in-depth exploration for the proposed covariance test procedure.

Our focus of this paper is the implementation of nonparametric covariance test for discretely observed functional data, supported by mathematically rigorous theory. To this end, we first consider a non-standardized statistic via sample-splitting technique \citep{Meinshausen09Pvaluehdflr} to overcome the dependence among estimates. Under the null hypothesis and certain regularity conditions,
we prove that 
the non-standardized statistic converges to a quadratic form of a Gaussian measure residing in infinite-dimensional space that depends on unknown parameters. Then, for practical use, we propose a normalized test statistic in the finite dimensions and perform Chi-squared test whose degrees of freedom is allowed to grow with the sample size. Importantly, we reveal the relationship between the permissible truncation level, sampling frequency and sample size, ensuring a reliable testing procedure with valid asymptotic null distributions. We stress that, even for the fully observed functional data, the asymptotic behaviour of a potentially diverging truncation is not well understood. Thus as a byproduct of our research, we also study the asymptotic behaviour of a non-standardized statistic in this ``ideal'' scenario, whose convergence depends on the diverging truncation level. This level should not exceed the order of $n^{1/(2\alpha+2)}$, where $\alpha>1$ is the polynomial decay rate of eigenvalues of the covariance operators, and lays a benchmark for analyzing the asymptotic behavior in the case of discretely observed data.

We summarize the main findings as follows. For the non-standardized statistic in \eqref{testdiscrete}, when the sampling frequency $N$ exceeds the order of $n^{1/(\alpha+1)}$, the maximum allowable range of the truncation level can nearly reach the order of $n^{1/(2\alpha+2)}$ with a quadratic form of a Gaussian measure, which coincides with the fully observed case. While for the normalized test statistic applied in practice, when $N$ exceeds the order of $n^{(\alpha+1)/(\alpha+3)}$ (viewed as the ``dense'' paradigm here), the allowable truncation level can be up to the order of $n^{1/(2\alpha+6)}$ for valid asymptotic Chi-squared null distributions, with a slight sacrifice of theoretical rate compared to $n^{1/(2\alpha+2)}$ for non-standardized statistic. It is worth noting that the required order of $N$ is higher than that in the mean and covariance estimation. This reflects the elevated difficulty arising from estimating diverging series of eigenfunctions involved in the inferential procedure. When $N$ is of a smaller order of $n^{(\alpha+1)/(\alpha+3)}$ (viewed as the ``sparse'' paradigm here), the allowable truncation level may still be close to the order of $N^{1/(2\alpha+2)}$, for which the asymptotic Chi-squared null distributions remain valid. Thus, in this relatively ``sparse'' paradigm, we recommend to shrink the truncation level appropriately but still allow it to slowly increase with the sample size $n$. Altogether, our investigation provides theoretical justification and practical guidance on when and how the functional covariance test procedure works in a nonparametric manner for discretely observed functional data that range from ``sparse'' to ``dense'' paradigms by allowing growing truncation level.

The remainder of this paper is organized as follows. In Section \ref{method}, we begin with reviewing and developing the covariance test for fully observed functional data, then proceed to describe the non-standardized statistic for discretely observed case. In Section \ref{theory}, we investigate the asymptotic behaviour of the non-standardized statistics, laying the foundation of mathematically rigorous theory for practically applied test statistic. In Section \ref{implementation},  we normalize the non-standardized statistic in finite dimensions and derive its asymptotic Chi-squared null distribution, revealing the connections between the truncation level, sampling frequency and sample size. Based on implemented methods provided in previous section, we perform the numerical studies through simulated and real data examples in Section \ref{simulation} and Section \ref{realdata}, respectively, and a conclusion is provided in Section \ref{discussion}. The technical proofs and additional discussions and simulations are deferred to Supplementary Material.


\section{Covariance Test from Continum to Discretized Observations}\label{method}

We start with two square integrable stochastic processes $X(s)$ and $Y(t)$ both defined on the interval $[0,1]$. For simplicity, we assume that the mean functions $\E[X(s)]=0$ and $\E[Y(t)]=0$, as the focus is the comparison of the second-order structures. Let $\mathcal{C}_X(s,s')=\cov\{X(s),X(s')\}=\E[X(s)X(s')]$ be the covariance function of the process $X(s)$. With abuse of notation,  $\mathcal{C}_X$ also denotes the integral operator associated with $\mathcal{C}_X(s,s')$, i.e., the covariance operator.
Similarly, for $Y(t)$, we analogously define $\mathcal{C}_Y$. We would like to test the following hypothesis
\begin{equation} \label{formula}
H_0: \mathcal{C}_X=\mathcal{C}_Y\qquad \text{v.s.}\qquad H_A: \mathcal{C}_X\neq \mathcal{C}_Y.
\end{equation}
This is intrinsically a nonparametric problem due to the infinite-dimensional nature, and can be equivalently formulated by making use of various distances \citep{Pigoli14distancecov}. The commonly adopted one is the Hilbert-Schmidt norm, denoted by $\|\cdot\|_{HS}$, to measure the difference between two covariance operators. Then the problem \eqref{formula} is equivalent to testing
\begin{equation} \label{formula:norm}
H_0: \left\|\mathcal{C}_X-\mathcal{C}_Y\right\|_{HS}=0\qquad \text{v.s.}\qquad H_A: \left\|\mathcal{C}_X-\mathcal{C}_Y\right\|_{HS}>0.
\end{equation}

In this paper we focus on developing a consistently nonparametric test for problem \eqref{formula:norm} when $X(s)$ and $Y(t)$ are discretely observed, supported by mathematically rigorous theory and practically effective implementation.
To this end, we first consider a non-standardized statistic and study its asymptotic behaviour under the infinite-dimensional framework with the help of infinite-dimensional measure. Then, for practical use, we propose a normalized test statistic in the finite dimensions and perform Chi-squared test whose degrees of freedom is allowed to grow with the sample size.
In the following, we begin with a review of existing works on fully observed functional data. Furthermore, we propose a non-standardized statistic in this continuous scenario, which provides a theoretical benchmark for asymptotic analysis on discretely observed data. Then we introduce the non-standardized statistic in the discretely observed scenario.


\subsection{Preliminary on covariance test for fully observed functions} \label{fullydata}

Assume that we have two groups of fully observed functions $X_i(s)$ ($1\leq i\leq n$) and $Y_i(t)$ ($1\leq i\leq m$) that are independent and identically distributed (i.i.d.) copies of $X(s)$ and $Y(t)$, respectively. Let $\widetilde{\mathcal{C}}_X(s,s')=n^{-1}\sum_{i=1}^n\{X_i(s)-\bar{X}(s)\}\{X_i(s')-\bar{X}(s')\}$ be the sample covariance function of $X(s)$, where $\bar{X}(s)=n^{-1}\sum_{i=1}^nX_i(s)$ is the sample mean. The analogues for $Y(t)$ is defined as $\widetilde{\mathcal{C}}_Y(t,t')$.
The fully-functional test statistic is constructed by plugging in the empirical version of covariances into \eqref{formula:norm} as $(n+m)\|\widetilde{\mathcal{C}}_X-\widetilde{\mathcal{C}}_Y\|_{HS}$, whose asymptotic null distribution depends on unknown infinite-dimensional parameters. Hence, critical values are approximated by some computationally intensive resampling methods such as bootstrap \citep{Paparoditis16bootstrap,Boente18multiplecovtest} and permutation \citep{Pigoli14distancecov}.

Besides the ``plug-in and resampling'' methods, the FPC-based test statistics are widely considered \citep{Panaretos10eqaulcov,Kraus12,Horvath13equalcov}. Let $\{(\vartheta_j,e_j)\}_{j=1}^\infty$ be the eigenvalues and eigenfunctions associated with the pooled covariance operator defined by $\mathcal{C}_P=\rho\,\mathcal{C}_X+(1-\rho)\mathcal{C}_Y$, where we assume that $n/(n+m)\rightarrow \rho$ for a fixed $\rho\in(0,1)$ as $n, m\rightarrow\infty$. The eigenfunctions $\{e_j\}_{j=1}^\infty$ form a complete orthonormal basis in $L^2[0,1]$ \citep{Hall06eigen,Hall06dispca}, so we can write $\|\mathcal{C}_X-\mathcal{C}_Y\|_{HS}^2=\sum_{j,k=1}^\infty\langle(\mathcal{C}_X-\mathcal{C}_Y)e_j,e_k\rangle^2$, where $\langle\cdot,\cdot\rangle$ denotes the inner product in $L^2[0,1]$. Given a truncation parameter,~\cite{Panaretos10eqaulcov} proposed an FPC-based test statistic tailored to the Gaussian assumption, while \cite{Horvath13equalcov} extended their work to the non-Gaussian case. Additionally, \cite{Kraus12} proposed an FPC-based test for comparing the dispersion operators, which is resistant to atypical observations and departures from Gaussian assumption. All of them demonstrated that the proposed test statistics converge to Chi-squared null distributions.

By comparison of existing works, we find that the derived asymptotic distributions of FPC-based test statistics facilitated by dimension reduction are more tractable, and this contributes to revealing the connections between theoretical principles and practical implementations. Under this consideration, we adopt the FPC-based framework for developing covariance test on discretely observed functional data.
It is important to note that, the existing FPC-based covariance tests are merely justified for fully observed functions with {\em fixed} truncation that does not grow with the sample size \citep{Panaretos10eqaulcov,Horvath13equalcov}. This may incur the risk of not capturing the difference that lies in high-order components. To circumvent this potential risk, we shall develop a consistently nonparametric test by letting the truncation grow in a suitable order with the sample size.
As a benchmark, we propose a non-standardized statistic on fully observed functional data in the following, and leave the discretely observed case to the subsequent subsection.

Let $\{(\tilde{\vartheta}_j,\tilde{e}_j)\}_{j=1}^K$ be the top $K$ pairs of eigenvalues and eigenfunctions associated with the empirical pooled covariance operator $\widetilde{\mathcal{C}}_P=(n+m)^{-1}(n\widetilde{\mathcal{C}}_X+m\widetilde{\mathcal{C}}_Y)$. We propose the following statistic
\begin{equation} \label{testfull}
\widetilde{T}_{K}
=\frac{nm}{n+m}\sum_{1\leq j\leq k\leq K}
\left\langle\left(\widetilde{\mathcal{C}}_X-\widetilde{\mathcal{C}}_Y\right)\tilde{e}_j,\tilde{e}_k\right\rangle^2.
\end{equation}
In Section \ref{theory}, we derive the asymptotic null distribution of this non-standardized statistic under the regime, where the truncation level $K$ grows with the sample size $n$. This theoretical result addresses a key gap in existing FPC-based testing approaches, and lays a foundation for studying the asymptotic behavior of covariance tests in the context of discretely observed functional data.

\subsection{Covariance test for discretely observed data}\label{discretelydata}

In practice, one does not observe the true functions $X_i(s)$ and $Y_i(t)$, but instead the discretized observations with noise contamination, i.e.,
\begin{equation}\label{observation}
\begin{split}
&X_{ip}=X_i(s_{ip})+\epsilon_{ip},\qquad p=1,\ldots,N_i,\quad i=1,\ldots,n,\\
&Y_{iq}=Y_i(t_{iq})+\varepsilon_{iq},\qquad q=1,\ldots,M_i,\quad i=1,\ldots,m,
\end{split}
\end{equation}
where, for simplicity, the measurement errors $\epsilon_{ip}$ and $\varepsilon_{iq}$ are assumed i.i.d. copies of $\epsilon$ and $\varepsilon$, respectively, and the observed times $s_{ip}$ and $t_{iq}$ are i.i.d. copies of $T$ following the uniform distribution $\mathcal{U}[0,1]$. The discussion on general sampling schemes is deferred to Supplementary Material. We also assume that $X, Y, \epsilon, \varepsilon$ and $T$ are mutually independent, and the sampling frequencies $N_i=N$ and $M_i=M$ for technical convenience in the sequel.

We stress that, when dealing with discretely observed data, the perturbation bounds \citep{Hall07flr} based on the cross-sectional sample covariance are no longer valid. The challenges arise due to pool-smoothing estimation based on discretely observed data with noise, and are further aggravated when considering the diverging index in a nonparametric fashion. With the aim to gain a comprehensive insight, we propose a non-standardized statistic of the FPC-based type and its normalized version that is implemented in practice is introduced in Section \ref{implementation}.

We employ the kernel smoothing by pooling all observations together for the estimation of covariance functions. Denote $G_{X,i}(s_{ip},s_{ip'})=X_{ip}X_{ip'}$ and define the kernel smoother by
\begin{equation}\label{cov:estimate}
\widehat{\mathcal{C}}_X(s,s')
=\frac{1}{nN(N-1)h^2}\sum_{i=1}^n\sum_{p\neq p'}^N\mathcal{K}\left(\frac{s_{ip}-s}{h}\right)\mathcal{K}\left(\frac{s_{ip'}-s'}{h}\right)G_{X,i}(s_{ip},s_{ip'}),
\end{equation}
where $\mathcal{K}$ is a bounded symmetric density kernel, and $\widehat{\mathcal{C}}_Y(t,t')$ is defined analogously. The choices of function $\mathcal{K}$ and the bandwidth $h$ are allowed to be different for $X$ and $Y$. We mention that the kernel smoother is adopted for technical convenience, while high-order choices, e.g., the local linear smoother \citep{Yao05fda,Li10ucr,Zhang16sparse}, are feasible.
To combine data appropriately, we choose the eigenfunctions associated with the pooled estimated covariance
\begin{equation} \label{poolcov:estimate}
\widehat{\mathcal{C}}_P(s,t)
=(n+m)^{-1}\left\{n\widehat{\mathcal{C}}_X(s,t)+m\widehat{\mathcal{C}}_Y(s,t)\right\},
\end{equation}
for $s,t\in[0,1]$. 
Then the empirical basis is defined by the following eigen-decomposition
\begin{equation} \label{eigenfun:estimate}
\int_{[0,1]}\widehat{\mathcal{C}}_P(s,t)\hat{e}_j(t)dt
=\hat{\vartheta}_j\hat{e}_j(s),\qquad 1\leq j\leq K,
\end{equation}
where $\hat{e}_j$ is the $j$-th eigenfunction of the pooled covariance $\widehat{\mathcal{C}}_P$ corresponding to the eigenvalue $\hat{\vartheta}_j$.
We emphasize that it is mathematically nontrivial to derive the asymptotic distribution of test statistic by directly plugging in the estimates $\widehat{\mathcal{C}}_X$, $\widehat{\mathcal{C}}_Y$ and $\hat{e}_j$'s into \eqref{testfull}, due to the complex structures of the kernel smoothers $\widehat{\mathcal{C}}_X$ and $\widehat{\mathcal{C}}_Y$ as well as the estimated eigenfunctions $\hat{e}_j$'s with diverging indices.

To this end, we note that the projections of the difference of covariance operators can be represented as $\langle(\mathcal{C}_X-\mathcal{C}_Y)e_j, e_k\rangle=\E(\xi_j\xi_k)-\E(\eta_j\eta_k)$, where $\xi_j$ and $\eta_j$ are the Fourier coefficients of $X$ and $Y$ projected onto the eigenfunction $e_j$ associated with the pooled covariance, respectively. This motivates us to consider the estimation of coefficients, and their second moments $\theta_{jk}:=\E(\xi_j\xi_k)$ and $\zeta_{jk}:=\E(\eta_j\eta_k)$, respectively. We adopt the Monte-Carlo (MC) average to approximate the integral due to its theoretical and practical convenience. For $1\leq i\leq n$ and $1\leq j\leq K$, the empirical MC-coefficients are defined as
\begin{equation} \label{score:estimate}
\begin{split}
\hat{\xi}_{ij}
=\frac{1}{N}\sum_{p=1}^{N}X_{ip}\hat{e}_j(s_{ip}),\qquad\qquad
\hat{\eta}_{ij}
=\frac{1}{M}\sum_{q=1}^{M}Y_{iq}\hat{e}_j(t_{iq}).
\end{split}
\end{equation}
For the second moment, it is instructive to note that $\E(\hat{\xi}_{ij}\hat{\xi}_{ik})\approx \E(\langle X_i,\hat{e}_j\rangle\langle X_i,\hat{e}_k\rangle)+O(1/N)$, where the asymptotic discretization error $O(1/N)$ arises from the diagonals $X^2_{ip}\hat{e}_j(s_{ip})\hat{e}_k(s_{ip})$ disturbed by the measurement errors $\epsilon_{ip}$ ($1\leq p\leq N$). On account of this, we estimate the second moment for individual subject by removing the diagonal terms,
\begin{equation} \label{eachcovar:estimate}
\begin{split}
&\hat{\theta}_{jk,i}
=\frac{1}{N(N-1)}\sum^N_{p\neq p'}X_{ip}X_{ip'}\hat{e}_j(s_{ip})\hat{e}_k(s_{ip'}),\\
&\hat{\zeta}_{jk,i}
=\frac{1}{M(M-1)}\sum^M_{q\neq q'}Y_{iq}Y_{iq'}\hat{e}_j(t_{iq})\hat{e}_k(t_{iq'}),
\end{split}
\end{equation}
and further define the estimates for the second moment by
\begin{equation} \label{covar:estimate}
\begin{split}
\hat{\theta}_{jk}
=\bar{\hat{\theta}}_{jk}-\bar{\hat{\xi}}_j\bar{\hat{\xi}}_k,\qquad
\hat{\zeta}_{jk}
=\bar{\hat{\zeta}}_{jk}-\bar{\hat{\eta}}_j\bar{\hat{\eta}}_k,
\end{split}
\end{equation}
where $\bar{\hat{\theta}}_{jk}$ is the empirical average of $\hat{\theta}_{jk,i}$ ($1\leq i\leq n$), and analogously for $\bar{\hat{\zeta}}_{jk}$, $\bar{\hat{\xi}}_j$ and $\bar{\hat{\eta}}_j$.

Note that the subsequent theoretical analysis for above estimates encounters challenges arising from the dependence between the observed data $X_{ip}$ ($1\leq p\leq N, 1\leq i\leq n$), $Y_{iq}$ ($1\leq q\leq M, 1\leq i\leq m$) and estimated eigenfunctions $\hat{e}_j$ ($1\leq j\leq K$). 
In order to circumvent these difficulty and make the error quantification mathematically tractable, we employ a sample-splitting strategy, which has a long history and is widely used in various applications \cite[][among others]{Meinshausen09Pvaluehdflr,Wasserman09HDVS}. See \cite{Romano19splittingtest} for a review.
Specifically, without loss of generality, assume that $n=2n_0$, $m=2m_0$ and denote by
\begin{equation*}
\mathbf{X}
=\left\{X_{2i}: i=1,\ldots,n_0\right\},\qquad \mathbf{X}'=\left\{X_{2i-1}: i=1,\ldots,n_0\right\},
\end{equation*}
the two disjoint subsets on the first group, respectively, and $\mathbf{Y}$ and $\mathbf{Y}'$ with the same definitions on the second group. Then on the first split sample $\mathbf{Z}=\mathbf{X}\cup\mathbf{Y}$, we attain the estimates of eigenfunctions $\{\hat{e}_{j,\mathbf{Z}}\}_{j=1}^K$ in the same manner as \eqref{cov:estimate}--\eqref{eigenfun:estimate}. Following \eqref{score:estimate}--\eqref{covar:estimate} on the second split sample $\mathbf{Z}'=\mathbf{X}'\cup\mathbf{Y}'$, we define the empirical MC-coefficients
\begin{equation*}
\hat{\xi}_{ij,\mathbf{Z}'}=\frac{1}{N}\sum_{p=1}^{N}X_{ip}\hat{e}_{j,\mathbf{Z}}(s_{ip}),\qquad \hat{\eta}_{ij,\mathbf{Z}'}=\frac{1}{M}\sum_{q=1}^{M}Y_{iq}\hat{e}_{j,\mathbf{Z}}(t_{iq}),\qquad X_{ip}, Y_{iq}\in\mathbf{Z}',
\end{equation*}
and thus the estimates of their second moments $\hat{\theta}_{jk,\mathbf{Z}'}$ and $\hat{\zeta}_{jk,\mathbf{Z}'}$.
As a result, we propose the following statistic
\begin{equation} \label{testdiscrete}
\widehat{T}_{K,\mathbf{Z}'}
=\frac{n_0m_0}{n_0+m_0}\sum_{1\leq j\leq k\leq K}
\left(\hat{\theta}_{jk,\mathbf{Z}'}-\hat{\zeta}_{jk,\mathbf{Z}'}\right)^2.
\end{equation}
Taking turns reversing the roles of $\mathbf{Z}$ and $\mathbf{Z}'$, we can also define the statistic $\widehat{T}_{K,\mathbf{Z}}$.

Unlike existing standardized FPC-based statistics \citep{Panaretos10eqaulcov,Horvath13equalcov}, the statistic $\widehat{T}_{K,\mathbf{Z}'}$ is non-standardized and requires normalization for practical use. However, analyzing this non-standardized form offers two advantages. First, the statistic $\widehat{T}_{K,\mathbf{Z}'}$ more directly captures the intrinsically asymptotic behaviour of FPC-based covariance tests as the truncation $K$ grows with the sample size $n$. Second, asymptotic analysis of the non-standardized statistic reduces theoretical complexities arising from discretized observations with contaminated errors.
We prove in Section \ref{th:discretelydata} that $\widehat{T}_{K,\mathbf{Z}'}$ converges to a quadratic form of a Gaussian measure residing in infinite-dimensional space, which provides a theoretical foundation for implementing $\chi^2$ test in Section \ref{implementation}.

\section{Asymptotic Properties and Phase Transition}\label{theory}

In this section, we derive the asymptotic null distributions of the statistics in \eqref{testfull} and \eqref{testdiscrete} under the infinite-dimensional framework, in the sense that the truncation level $K$ grows with the sample size $n$ and $m$, and for \eqref{testdiscrete} we reveal a phenomenon of phase transition as the sampling frequencies $N$ and $M$ increase from ``sparse'' to ``dense'' paradigms. Without loss of generality, let $N\asymp M$ and $n\asymp m$.
We establish that the asymptotic null distribution based on discretely observed functional data remains valid for different allowable ranges of $K$ and, when $N$ reaches certain magnitude of $n$, it behaves as if the functions are fully observed. When $N$ is below this transition magnitude, we may shrink the allowable range of $K$ appropriately (still allowed to diverge with $n$) for which the asymptotic null distribution remains valid.

To properly characterize the asymptotic distribution in infinite-dimensional space, we introduce some notations below. By Mercer's theorem, the covariance functions admit the spectral decompositions $\mathcal{C}_X(s,s')=\sum_{j=1}^\infty\lambda_j\phi_j(s)\phi_j(s')$ and $\mathcal{C}_Y(t,t')=\sum_{j=1}^\infty\kappa_j\psi_j(t)\psi_j(t')$, and the corresponding processes admit the so-called Karhunen-Lo\`eve expansions $X(s)=\sum_{j=1}^\infty a_j\phi_j(s)$ and $Y(t)=\sum_{j=1}^\infty b_j\psi_j(t)$, where $a_j$ and $b_j$ are principal scores of $X$ and $Y$, respectively. Given any two elements $f, g\in L^2[0,1]$, denote their tensor product by $f\otimes g$ satisfying that $(f\otimes g)u=\langle f,u\rangle g$ for $u\in L^2[0,1]$. Let $\mathcal{B}_{HS}(L^2[0,1])$ be a Hilbert space equipped with Hilbert-Schmidt norm $\|\cdot\|_{HS}$ generated by the inner product $\langle\cdot,\cdot\rangle_{HS}$, which contains all Hilbert-Schmidt operators acting on $L^2[0,1]$. Similarly, denote the tensor product in $\mathcal{B}_{HS}(L^2[0,1])$ by $\otimes_{HS}$. For $j, k\in\mathbb{N}$, define the operator $\Phi_{jk}=e_j\otimes e_k$ that belongs to $\mathcal{B}_{HS}(L^2[0,1])$. It is easy to see that $\|\Phi_{jk}\|_{HS}=1$. With these notations, we define an operator on $\mathcal{B}_{HS}(L^2[0,1])$ that will be used in our theorems,
\begin{equation}\label{operator:M}
\mathscr{M}
=\sum_{j=1}^\infty\Phi_{jj}\otimes_{HS}\Phi_{jj}+\sum_{j\neq k}\frac{\left(\Phi_{jk}+\Phi_{kj}\right)\otimes_{HS}\left(\Phi_{jk}+\Phi_{kj}\right)}{8}.
\end{equation}
The operator $\mathscr{M}$ performs similar to the identity operator except for some coefficients on the orthonormal basis, since our statistics are constructed based on the upper-triangular elements (i.e., sum over $1\leq j\leq k\leq K$).

\subsection{Theoretical benchmark on fully observed functions}\label{th:fullydata}

In this subsection, we investigate the asymptotic null distribution of the statistic \eqref{testfull}, allowing the truncation level $K$ to grow with the sample size $n$. The derivation relies on the structure of cross-sectional sample covariances based on fully observed functions and the sharp perturbation bounds for estimated eigenfunctions with a diverging index \citep{Hall07flr}. A sharp upper bound for the maximum allowable $K$ for which the asymptotic null distribution remains valid is established, which sets up a stage for the justification of discretely observed data in the subsequent subsection.

\begin{assumption}\label{asmA1}
The processes satisfy $\sup_{s\in[0,1]}\E[X^8(s)]<\infty$ and $\sup_{t\in[0,1]}\E[Y^8(t)]<\infty$.
\end{assumption}
Note that the condition for controlling the fourth moment of processes is commonly used in the literature of FDA, e.g., covariance estimation \citep{Hall06dispca,Zhang16sparse} and testing with fixed truncation level \citep{Panaretos10eqaulcov,Horvath13equalcov}. However, when considering diverging truncation level, we need to control the eighth moment as in Assumption \ref{asmA1} to quantify the expectation of  high-order moment, reflecting the elevated theoretical complexity. This moment condition holds for Gaussian process with continuous sample path \citep{Landau70gaussseprocess}.

\begin{assumption}\label{asmA2}
For each $j\in\mathbb{N}$ and $r=2,3,4$, we have $(1+\const_0^{-1})\E^r(a_j^2)\leq\E(a_j^{2r})\leq \const_0\E^r(a_j^2)$, where $\const_0>2$ and does not depend on $j$. Moreover, the sequence of product scores $\{a_ja_k\}_{j,k}$ with $j,k\in\mathbb{N}$ are uncorrelated. Similar assumptions hold for scores $b_j$ with $j\in\mathbb{N}$.
\end{assumption}
Assumption \ref{asmA2} imposes regular conditions on the scores of processes. The condition that the high-order moment of scores are bounded by their second moments is standard in FDA \citep{Hall06preflr,Hall07flr,Kong16pflr}. The uncorrelatedness of the sequence of $\{b_jb_k\}_{j,k}$ with $j,k\in\mathbb{N}$ is weaker than the independence assumption \citep{Dai17Classifier} and holds for Gaussian process. A similar assumption of this uncorrelatedness also appears in \cite{Hall06preflr}.

\begin{assumption}\label{asmA3}
Under $H_0$, there exists a constant $\alpha>1$ such that for all $j\in\mathbb{N}$,
\begin{equation*}
\const^{-1}j^{-\alpha}\leq\lambda_j\leq\const j^{-\alpha},\qquad\lambda_j-\lambda_{j+1}\geq\const^{-1}j^{-\alpha-1},
\end{equation*}
where $\lambda_j$'s are eigenvalues of $\mathcal{C}_X$, and $\const$ denotes a generic constant that is finite and positive.
\end{assumption}
Assumption \ref{asmA3} imposes the standard condition of covariance smoothness in terms of a polynomial decay rate of eigenvalues, preventing the spacings between adjacent order statistics from being too small. This condition is imposed for studying the asymptotic null distribution of the proposed test. Assumption \ref{asmA3} is commonly used in FDA, including techniques such as functional principle component analysis \citep{Hall06eigen, Li10ucr} and functional regression \citep{Hall07flr,Cai12prediction,Dou12FLR}.

Denote $\widetilde{\Delta}_X=\widetilde{\mathcal{C}}_X-\mathcal{C}_X$ and $\widetilde{\Delta}_Y=\widetilde{\mathcal{C}}_Y-\mathcal{C}_Y$, respectively. For each $l\in\mathbb{N}$, define that
\begin{equation} \label{eigengap}
\varrho_{X,l}
=\min_{k\neq l}\left|\lambda_k-\lambda_l\right|,\qquad\varrho_{Y,l}=\min_{k\neq l}\left|\kappa_k-\kappa_l\right|,
\end{equation}
and further define the event set as
\begin{equation} \label{eventE}
\mathcal{E}_l(n,m)
=\left\{\max\left(\|\widetilde{\Delta}_X\|_{HS}, \|\widetilde{\Delta}_Y\|_{HS}\right)<\min(\varrho_{X,l},\varrho_{Y,l})/2\right\},
\end{equation}
which denotes the set of all realizations such that \eqref{eventE} holds for sample sizes $n$ and $m$ \citep{Hall07flr}. Intuitively, the set $\mathcal{E}_l$ contains all realizations such that the estimator $\hat{\lambda}_l$ falls in a circle centered at $\lambda_l$ with radius $\varrho_{X,l}/2$, and from which the $l$-th eigenfunction are well estimated. The statement ``on $\mathcal{E}_l$'' should be interpreted as stating that the obtained bounds are valid for all realizations for which \eqref{eventE} holds. When the eigenvalues exhibit a polynomial decay rate as in Assumption \ref{asmA3}, the events \eqref{eventE} hold with high probability uniformly for $1\leq l\leq K$, with $K$ satisfying the condition \eqref{K:fully} below.

\begin{theorem} \label{theorem:fully}
Let $X(\cdot)$ and $Y(\cdot)$ be square integrable processes on $[0,1]$ with mean zero. Suppose that Assumptions \ref{asmA1}--\ref{asmA3} are satisfied for given value $\alpha$, and the null hypothesis holds. If the truncation level satisfies
\begin{equation}\label{K:fully}
K=o\left(n^{1/(2\alpha+2)}\right),
\end{equation}
as $n\rightarrow\infty$, then the probability of event \eqref{eventE} satisfies that $\mathbb{P}(\mathcal{E}_K)\rightarrow1$. Moreover, on the event $\mathcal{E}_K$, the statistic in \eqref{testfull} satisfies that
\begin{equation}\label{convergence:fully}
\widetilde{T}_K\stackrel{d}{\longrightarrow}\left\langle\mathscr{M}\mathcal{G}, \mathcal{G}\right\rangle_{HS},
\end{equation}
where $\mathcal{G}$ is a Gaussian measure in $\mathcal{B}_{HS}(L^2[0,1])$ and $\mathscr{M}$ is defined in \eqref{operator:M}.
In addition, the quadratic form in \eqref{convergence:fully} has a representation as below
\begin{equation}\label{quadraticform}
\langle\mathscr{M}\mathcal{G}, \mathcal{G}\rangle_{HS} =\sum_{j=1}^\infty\left[(1-\rho)\E(a^4_j)+\rho\E(b^4_j)-\lambda_j^2\right]W_{jj}^2
+\sum_{j<k}\lambda_j\lambda_kW_{jk}^2,
\end{equation}
where $\{W_{jk}\}_{1\leq j\leq k<\infty}$ are i.i.d. random variables following standard normal distribution.
\end{theorem}

From Theorem \ref{theorem:fully}, we observe that the maximum allowable range of $K$ for which the asymptotic null distribution remains valid can be close to the order of $n^{1/(2\alpha+2)}$. This result is surprising at first glance since $n^{1/(2\alpha+2)}$ aligns with the bound derived in \eqref{eventE}, implying that as long as the eigenfunction's are well estimated, the asymptotic null distribution remains valid for non-standardized statistic projected on these estimated eigenfunctions. This sharp truncation order is reasonable, since the error of non-standardized statistic approximating to the null distribution is negligible compared to that in \eqref{eventE}, facilitated by the sharp perturbation bounds of estimated eigenfunctions with diverging index on fully observed functions.
The insights in Theorem \ref{theorem:fully} provide a comparable ground for investigating the statistic \eqref{testdiscrete} constructed on discretely observed data, and help determine the minimum sample frequency required to achieve the same asymptotic behaviour as if the functions were fully observed.

\subsection{Asymptotic property for discretely observed data}\label{th:discretelydata}

In this subsection we study the asymptotic behaviour of the proposed statistic \eqref{testdiscrete}, allowing the sampling frequency $N$ and the truncation level $K$ to grow with the sample size $n$. As discussed in Section \ref{th:fullydata}, the sharp truncation order of non-standardized statistic \eqref{testfull} is owed to the sharp perturbation bounds of estimated eigenfunctions with diverging index on fully observed functions. In the context of discretely observed functional data, we present a new result on the perturbation bounds of estimated eigenfunctions with diverging index based on pool-smoothing covariance estimation. As aforementioned, the existing perturbation bounds are for either fully observed curves or fixed index, until a recent work \citep{ZWY22flrdis} makes substantial progress. This enables us to explore the asymptotic behavior of the proposed statistic. Before stating the result, we introduce a new regularity condition \citep[see A.4 in][]{ZWY22flrdis} on the series of eigenfunctions.
\begin{assumption}\label{asmA4}
Assume that, for all $j\in \mathbb{N}$, $\sup_{s\in[0,1]}|\phi_j(s)|=O(1)$ and for $r=1,2$,
\begin{equation*}
\sup_{s\in[0,1]}|\phi^{(r)}_j(s)|\lesssim j^{\beta/2}\sup_{s\in[0,1]}|\phi^{(r-1)}_j(s)|,
\end{equation*}
where $\beta$ is a positive constant. The same assumptions hold for the eigenfunctions $\{\psi_j\}_{j\in \mathbb{N}}$.
\end{assumption}
Assumption \ref{asmA4} characterizes the frequency increment of each eigenfunction via the smoothness of its derivatives, and for some standard orthonormal basis, such as Fourier (used in our simulations), Legendre and wavelet, $\beta=2$. On the split sample $\mathbf{Z}$, denote $\widehat{\Delta}_{X, \mathbf{Z}}=\widehat{\mathcal{C}}_{X, \mathbf{Z}}-\mathcal{C}_X$ and $\widehat{\Delta}_{Y, \mathbf{Z}}=\widehat{\mathcal{C}}_{Y, \mathbf{Z}}-\mathcal{C}_Y$. For each $l\in\mathbb{N}$, similar to $\mathcal{E}_l$ in \eqref{eventE}, we define the event set
\begin{equation} \label{eventF}
\mathcal{F}_l(n,m,N,M,h)
=\left\{\max\left(\|\widehat{\Delta}_{X, \mathbf{Z}}\|_{HS}, \|\widehat{\Delta}_{Y, \mathbf{Z}}\|_{HS}\right)<\min(\varrho_{X,l},\varrho_{Y,l})/2\right\},
\end{equation}
where $\varrho_{X,l}$ and $\varrho_{Y,l}$ are defined in \eqref{eigengap}. It is seen that the event sets are nested for increasing $l\in\mathbb{N}$, saying $\mathcal{F}_l\subset\mathcal{F}_{l-1}\cdots\subset\mathcal{F}_1$, and hold with high probability uniformly for $1\leq l\leq K$ with $h$ and $K$ satisfying \eqref{bandwidth} and \eqref{K:discrete}, respectively. Slightly modifying the arguments in \cite{ZWY22flrdis} leads to the following proposition tailored for our problem.

\begin{proposition} \label{eigenfunction:prop}
Suppose that Assumptions \ref{asmA1}--\ref{asmA4} hold for given values $\alpha$ and $\beta$, and $K\in\mathbb{N}$ satisfies $h^4K^{2\alpha+2\beta}=O(1)$. Then under the null hypothesis and on the event $\mathcal{F}_K$, the estimated eigenfunctions in \eqref{eigenfun:estimate} satisfy that, as $n\rightarrow\infty$,
\begin{equation} \label{eigenfunction:bound}
\E\left(\|\hat{e}_{j,\mathbf{Z}}-e_j\|^2\right)\lesssim\frac{j^2}{n}\left(1+\frac{j^{2\alpha}}{N^2}\right)
+\frac{j^\alpha}{nNh}\left(1+\frac{j^\alpha}{N}\right)+h^4j^{2\beta+2},\qquad 1\leq j\leq K.
\end{equation}
\end{proposition}

Here the constraint $h^4K^{2\alpha+2\beta}=O(1)$ is to suppress the bias of the estimated eigenfunctions based on pool-smoothing estimation of discretely observed data. Note that the magnitude of the diverging truncation level $K$ is jointly determined by the observed data (e.g., $n,N$), the smoothing parameter $h$, the decay rate $\alpha$ of eigenvalues and the frequency increments rate $\beta$ of eigenfunctions. This proposition lays the foundation for asymptotic analysis of the proposed statistic \eqref{testdiscrete}. To comprehend this, for any fixed index $j$, the bound \eqref{eigenfunction:bound} can be simplified to $n^{-1}+(nN)^{-4/5}$ by taking $h=(nN)^{-1/5}$. Moreover, when $N\gtrsim n^{1/4}$, the bound \eqref{eigenfunction:bound} reaches the parametric rate $n^{-1}$ that agrees with those for the mean/covariance estimation \citep{Zhang16sparse} coupled with the classical perturbation bound \citep{Dauxois82fpca}. While $N=o(n^{1/4})$, this becomes a typical one-dimensional nonparametric rate $n^{-4/5}$ which is consistent with that in \cite{Hall06dispca}. When the index $j$ is allowed to grow with $n$ (specified by $\mathcal{F}_j$), and taking $\beta=2$ that holds for common basis, if $N \gtrsim n^{1/4+(\alpha-1)/(2\alpha+2)}$ the bound \eqref{eigenfunction:bound} can be reduced to $j^2/n$ which coincides with the optimal rate in the fully observed case \citep{Hall07flr}.

\begin{assumption}\label{asmA5}
Without loss of generality, the measurement errors $\epsilon$ and $\varepsilon$ both have mean zero and variance $\sigma^2<\infty$, and furthermore satisfy that $\E(\epsilon^4)<\infty$ and $\E(\varepsilon^4)<\infty$.
\end{assumption}

The Assumption \ref{asmA5} bounds the fourth moment of the measurement errors for discretely observed functional data. We mention that the errors $\epsilon$ and $\varepsilon$ are allowed to have different variances, and this has no impacts on our method and theory.

\begin{theorem} \label{theorem:discrete}
Let $X(\cdot)$ and $Y(\cdot)$ be square integrable processes on $[0,1]$ with mean zero. Suppose that Assumptions \ref{asmA1}--\ref{asmA5} are satisfied for given values $\alpha$ and $\beta$, and the null hypothesis holds. As $n\rightarrow\infty$, if the bandwidth satisfies
\begin{equation} \label{bandwidth}
N^{-1}n^{-\frac{1}{2}}\delta_{\text{min}}^{\frac{1}{2}}\lesssim h\lesssim \delta_{\text{max}}^{-\frac{1}{4}-\frac{\max\{\beta+1-\alpha,0\}}{4\alpha+4}},
\end{equation}
and the truncation level satisfies
\begin{equation} \label{K:discrete}
K=o\left(\delta_{\text{min}}^{1/(2\alpha+2)}\right),
\end{equation}
where $\delta_{\text{min}}=\min\{N^{\alpha+1}, n\}$ and $\delta_{\text{max}}=\max\{N^{\alpha+1}, n\}$, then the probability of event \eqref{eventF} satisfies that $\mathbb{P}(\mathcal{F}_K)\rightarrow1$. Moreover, on the event $\mathcal{F}_K$, the statistic \eqref{testdiscrete} satisfies that
\begin{equation} \label{convergence:discrete}
\widehat{T}_{K,\mathbf{Z}'}\stackrel{d}{\longrightarrow}\left\langle\mathscr{M}\mathcal{G}, \mathcal{G}\right\rangle_{HS},
\end{equation}
where $\mathcal{G}$ is a Gaussian measure in $\mathcal{B}_{HS}(L^2[0,1])$ and $\mathscr{M}$ is defined in \eqref{operator:M}. In addition, the quadratic form in \eqref{convergence:discrete} has the same representation as in \eqref{quadraticform} of Theorem \ref{theorem:fully}.
\end{theorem}

Theorem \ref{theorem:discrete} gives an insight between the truncation level $K$ and the sampling frequency $N$. When the number of observations $N$ exceeds the order of $n^{1/(\alpha+1)}$, one can see that the maximum allowable range of $K$ can be arbitrarily close to the magnitude of $n^{1/(2\alpha+2)}$, which coincides with the fully observed case as presented in Theorem \ref{theorem:fully}. On the other hand, when the number of observations $N$ is below the order of $n^{1/(\alpha+1)}$, according to \eqref{K:discrete} the maximum allowable range of $K$ becomes smaller, but still nearly reaches the order of $N^{1/2}$ that depends on the relatively ``sparse'' sampling frequency $N=o(n^{1/(\alpha+1)})$.

For the non-standardized statistic \eqref{testdiscrete}, the phenomenon of phase transition occurs at the order of $n^{1/(\alpha+1)}$, depending on the decay rate of eigenvalues associated with the covariance operator. Compared to the transition order of $n^{1/4}$ in terms of mean/covariance estimation via pool-smoothing method, 
this crossover effect on $\alpha$ reflects the intrinsic complexity in covariance test which has a close relation to the covariance smoothness. When considering the normalization of statistic \eqref{testdiscrete} for practical use, the estimation of variances whose magnitude inherently link to the covariance smoothness introduces additional challenges in covariance test, and these difficulties are further elevated for growing number of estimated variances and discretely observed functional data. 

\section{The  Chi-squared Test and Implementation}\label{implementation}

In Section \ref{theory} we prove that when the sample size and the truncation level both grow up, the non-standardized statistic $\widehat{T}_{K,\mathbf{Z}'}$ converges weakly to a null distribution characterized by an infinite-dimensional series that depends on unknown parameters. The infinite-dimensional series is a quadratic form of a Gaussian measure residing in infinite-dimensional space. For practical implementation, due to finite sample size we must project Gaussian measure onto a finite-dimensional subspace (i.e., finite truncation $K$). In such case, we can standardizing the statistic $\widehat{T}_{K,\mathbf{Z}'}$ by estimating unknown parameters and obtain the asymptotic Chi-squared null distribution with degrees of freedom $K(K+1)/2$. We stress that the projected dimension $K$, though finite, is allowed to increase with the sample size, different from the fixed $K$ in existing works \citep{Panaretos10eqaulcov,Horvath13equalcov}.

In the remaining of this section, we propose a normalized version of $\widehat{T}_{K,\mathbf{Z}'}$ for a given truncation $K$, and derive its asymptotic null distribution under the framework of increasing $K$. Furthermore, by examining the allowable range of $K$ for which the asymptotic null distribution remains valid, a phenomenon of phase transition is revealed. This provides theoretical justification and practical guidance for implementing test procedure.

To be specific, according to the proof of Theorem \ref{theorem:fully} in Supplementary Material, under $H_0$ we obtain the asymptotic variances of $\hat{\theta}_{jk}-\hat{\zeta}_{jk}$ ($1\leq j\leq k\leq K$) defined as
\begin{equation*}
\rho_{jk}
=(1-\rho)\left[\E(a_j^2a_k^2)-\E^2(a_ja_k)\right]+\rho\left[\E(b_j^2b_k^2)-\E^2(b_jb_k)\right].
\end{equation*}
Under Assumption \ref{asmA2}, the variances reduce to the corresponding weights in \eqref{quadraticform}, i.e., $(1-\rho)\E(a^4_j)+\rho\E(b^4_j)-\lambda_j^2$ for $j=k$ and $\lambda_j\lambda_k$ for $j\neq k$. Given the estimated eigenfunctions $\hat{e}_{j,\mathbf{Z}}$ ($1\leq j\leq K$) on the first split sample $\mathbf{Z}$, define the empirical MC-coefficient as in \eqref{eachcovar:estimate} and its empirical average on the second split sample $\mathbf{Z}'$. Then we define the estimates for the variances as
\begin{equation} \label{var:estimate}
\hat{\rho}_{jk,\mathbf{Z}'}
=\frac{m_0}{n_0+m_0}\left\{\frac{1}{n_0}\sum_{i=1}^{n_0}\hat{\theta}^2_{jk,i}
-\bar{\hat{\theta}}_{jk}^2\right\}
+\frac{n_0}{n_0+m_0}\left\{\frac{1}{m_0}\sum_{i=1}^{m_0}\hat{\zeta}^2_{jk,i}-\bar{\hat{\zeta}}_{jk}^2\right\}.
\end{equation}
Consequently, we propose the following standardized test statistic
\begin{equation} \label{testdiscrete:normalize}
\widehat{D}_{K,\mathbf{Z}'}
=\frac{n_0m_0}{n_0+m_0}\sum_{1\leq j\leq k\leq K}
\frac{(\hat{\theta}_{jk,\mathbf{Z}'}-\hat{\zeta}_{jk,\mathbf{Z}'})^2}{\hat{\rho}_{jk,\mathbf{Z}'}}.
\end{equation}

For fully observed functions, the fourth moment is also utilized for standardization by \cite{Horvath13equalcov}, while \cite{Panaretos10eqaulcov} use the product of second moments under the Gaussian assumption. In our extensive numerical comparisons, the estimates for variances based on fourth moment exhibit the stable performance due to diagonal-elimination for second moment estimation.  Although the existing FPC-based test statistics share similar types to the normalized statistic $\widehat{D}_{K,\mathbf{Z}'}$, they perform poorly on the sparse functional data because of non-negligible smoothing errors. The utilization of pool-smoothing method could improve their performance. In Section S.5 of Supplementary Material, we provide simulated studies and detailed discussion on this issue, and additionally we investigate the performance of \eqref{testdiscrete:normalize} by replacing \eqref{var:estimate} with the product of second moment estimation.



Similar to event set \eqref{eventE}, for each $l\in\mathbb{N}$ we define an event set containing all realizations for well-estimated variances as below
\begin{equation} \label{eventH}
\mathcal{H}_l(n,m,N,M,h)
=\left\{|\hat{\rho}_{jk,\mathbf{Z}'}-\rho_{jk}|<\min(\lambda_j\lambda_k,\kappa_j\kappa_k)
/(2\const_0),~\forall~ 1\leq j,k\leq l\right\},
\end{equation}
where $\const_0$ is the constant defined in Assumption \ref{asmA2}. This event holds with high probability uniformly for $1\leq l\leq K$ as long as $h$ and $K$ satisfy \eqref{bandwidth:normalize} and \eqref{K:discrete-normalize}, respectively. To properly describe the asymptotic distribution of $\widehat{D}_{K,\mathbf{Z}'}$ with the index $K$ allowed to diverge with the sample size $n$, we introduce the following Prokhorov metric, which provides a quantitative measure for metrizing the weak convergence of measures \citep{Gibbs02metric},
\begin{equation*}
\pi(\mu,\nu):=\inf\left\{\delta >0: \mu(B)\leq\nu(B^\delta)+\delta~~\text{for~any~}B\in\mathcal{B}\right\},
\end{equation*}
where $\mu$ and $\nu$ are two probability measures on a metric space $(\Omega, d)$ equipped with a Borel $\sigma$-algebra $\mathcal{B}$ and $B^\delta:=\{x: \inf_{y\in B}d(x,y)\leq\delta\}$.

\begin{theorem} \label{theorem:discrete-normalize}
Let $X(\cdot)$ and $Y(\cdot)$ be square integrable processes on $[0,1]$ with mean zero. Suppose that Assumptions \ref{asmA1}--\ref{asmA5} are satisfied for given values $\alpha$ and $\beta$, and the null hypothesis holds.
As $n\rightarrow\infty$, if the bandwidth satisfies
\begin{equation} \label{bandwidth:normalize}
(nN)^{-1}\tilde{\delta}_{\text{min}}^{\frac{3\alpha+4}{2\alpha+6}}\lesssim h\lesssim \tilde{\delta}_{\text{max}}^{-\frac{1}{4}-\frac{\beta}{4\alpha+12}},
\end{equation}
and the truncation level satisfies
\begin{equation} \label{K:discrete-normalize}
K=o\left(\tilde{\delta}_{\text{min}}^{1/(2\alpha+6)}\right),
\end{equation}
where $\tilde{\delta}_{\text{min}}=\min\{N^{(\alpha+3)/(\alpha+1)}, n\}$ and $\tilde{\delta}_{\text{max}}=\max\{N^{(\alpha+3)/(\alpha+1)}, n\}$, then the probability of the intersection of events \eqref{eventF} and \eqref{eventH} satisfies that $\mathbb{P}(\mathcal{F}_K\cap\mathcal{H}_K)\rightarrow1$. Furthermore, on the event $\mathcal{F}_K\cap\mathcal{H}_K$, it holds that
\begin{equation} \label{convergence:discrete-normalize}
\pi\left(\widehat{D}_{K,\mathbf{Z}'},\chi^2_{K(K+1)/2}\right)\longrightarrow0,
\end{equation}
where $\pi(\cdot,\cdot)$ is the Prokhorov metric, and $\chi^2_{K(K+1)/2}$ is a Chi-squared distribution with degrees of freedom $K(K+1)/2$ that is allowed to diverge with the sample size $n$.
\end{theorem}

The result in Theorem \ref{theorem:discrete-normalize} is indeed consistent with our intuition that more observations per subject allow the test statistic to effectively detect differences in more components along the spectra of covariance operators. This fact can be confirmed by \eqref{K:discrete-normalize}, exhibiting a different phenomenon of phase transition from the non-standardized case in Theorem \ref{theorem:discrete}. Specifically, when the number of observations $N$ exceeds the order of $n^{(\alpha+1)/(\alpha+3)}$ (viewed as the ``dense'' paradigm for covariance test), the maximum allowable range of $K$ can be arbitrarily close to the magnitude of $n^{1/(2\alpha+6)}$; while the number of observations is below the order of $n^{(\alpha+1)/(\alpha+3)}$ (the ``sparse'' paradigm), the maximum allowable range shrinks to the order of $N^{1/(2\alpha+2)}$; both of them are slower than $n^{1/(2\alpha+2)}$ in the non-standardized case. Moreover, the transition order $n^{(\alpha+1)/(\alpha+3)}$ with $\alpha>1$ in Assumption \ref{asmA3} is larger than the order of $n^{1/4}$ in covariance estimation, underscoring the relatively higher complexity of covariance test.

The deterioration of rates from \eqref{K:discrete} to \eqref{K:discrete-normalize} reflects elevated difficulty in performing covariance test with tractable distribution due to estimating a diverging number of variances that converge to zero as $K$ increases to infinity. This sacrifice of allowable range is reasonable since the inflated order consists of three components: the sum of double indices $K^2$ in the test statistic, the sharp error bounds $K/\sqrt{n}$ for diverging number of estimated eigenfunctions, and the instability of inverse of small variances $\rho_{jk}\asymp K^{2\alpha}$. We find that \cite{Panaretos10eqaulcov} have made an attempt to construct a covariance test with diverging truncation on fully observed functional data in their supplementary material. However, they applied classical perturbation bounds of estimated eigenfunctions with diverging indices and obtained a rough truncation order of $n^{1/(3\alpha+14)}$, which is much slower than ours.

Theorem \ref{theorem:discrete-normalize} provides theoretical foundation and practical guidance for applications of standardized statistic $\widehat{D}_{K,\mathbf{Z}'}$. According to \eqref{convergence:discrete-normalize}, we can perform Chi-squared tests based on $\widehat{D}_{K,\mathbf{Z}'}$ for a range of $K$. Apparently $\widehat{D}_{K,\mathbf{Z}}$ can be defined similarly by taking turns reversing the split samples and have the same property as $\widehat{D}_{K,\mathbf{Z}'}$. Therefore, we take into account pooling two statistics together to calculate the rejection rate in simulation, while taking average of two $p$-values in application. In addition, we can randomly conduct equal-splitting and repeat this procedure many times on whole data to obtain a collection of test statistics and then combine them following the same way.

In practice, the number of projections considered is important for the FPC-based covariance test. The existing selection rules are mainly designed for fully observed functional data and based on the complexity of the whole samples \citep{Panaretos10eqaulcov,Horvath13equalcov}. These rules lack theoretical guarantees and are not applicable to discretely observed data. The condition \eqref{K:discrete-normalize} provides theoretical justification for the maximum allowable range of $K$ (depend on $n$ and $N$) that cannot be too large for validity of the test. On the other hand, one might hope to include as many components as possible to not miss important signals. Given such considerations, instead of providing a truncation selection rule, we suggest an effective implementation particularly for the test based on our theoretical findings, which investigates the performance by running the test over a suitable range of different $K$ and makes a comprehensive and valid justification by the trend of testing performance. It can be checked that the range of bandwidth in \eqref{bandwidth} indeed covers the bandwidth choices for covariance estimation in both ``sparse'' and ``dense'' cases of \cite{Zhang16sparse}. This suggests that our testing procedure based on pool-smoothing estimation is not very sensitive to the choice of $h$, i.e., allows for a range of bandwidth values.

\section{Simulation}\label{simulation}
To assess the performance of the proposed covariance test \eqref{testdiscrete:normalize}, denoted by $T_{pool}$, we carry out simulation studies by comparison with existing methods including $T_{PK}$ in \cite{Panaretos10eqaulcov} and $T_{FH}$ in \cite{Horvath13equalcov}, where the later use $p$ as the truncation. 
We generate the measurements $X_{ip}=X_i(s_{ip})+\epsilon_{ip}$ for $i=1,\ldots,n$ and $p=1,\ldots,N$, where the observed times $s_{ip}$ are i.i.d. from the uniform distribution $\mathcal{U}[0,1]$ and independent of $\epsilon_{ip}\sim\mathcal{N}(0,0.1^2)$. The underlying trajectories $X_i(s)=\sum_{j=1}^{50}\sqrt{\lambda_j}a_{ij}\phi_j(s)$ and i.i.d. scores $a_{ij}\sim\mathcal{N}(0,1)$ are independent of $s_{ip}$ and $\epsilon_{ip}$. The eigenfunctions are $\phi_j(s)=\sqrt{2}\sin(j\pi s)$ and the eigenvalues are $\lambda_1=1, \lambda_2=0.64$ and $\lambda_j=j^{-1.5}$ with $3\leq j\leq 50$. It is easily checked that the triple parameters $(a_{ij},\lambda_j,\phi_j)$ and the process $X_i(s)$ satisfy Assumptions \ref{asmA1}--\ref{asmA4}, as well as the measurement errors $\epsilon_{ip}$ satisfy Assumption \ref{asmA5}. The second group $Y_{iq}=Y_i(t_{iq})+\varepsilon_{iq}$ for $i=1,\ldots,m$ and $q=1,\ldots,M$ are independently generated from the underlying trajectories $Y_i(t)=\sum_{j=1}^{50}(1+\gamma_j)\sqrt{\lambda_j}b_{ij}\phi_j(t)$, where the scaling vector $\mathbb{\gamma}=(\gamma_1,\ldots,\gamma_{50})^\top$ measures the deviation from the null hypothesis corresponding to $\gamma=\mathbf{0}$, and the observed times $t_{iq}$, the error $\varepsilon_{iq}$ and the scores $b_{ij}$ are generated from the same setting as $s_{ip}, \epsilon_{ip}, a_{ij}$, respectively.

To evaluate the performance of the proposed tests under the null hypothesis and against various alternatives, we consider the following two scenarios:
\begin{itemize}
	\item Scenario I~~: $\gamma=a(\mathbf{e}_1+\mathbf{e}_2)$, where $a$ increases from $0$ to $0.6$;
	\item Scenario II~: $\gamma=a(\mathbf{e}_3+\mathbf{e}_4)$, where $a$ increases from $0$ to $0.6$.
\end{itemize}
Here $\mathbf{e}_k$ denotes the standard basis vector with $1$ in the $k$-th coordinate and $0$ elsewhere, and $a$ represents the signal strength. In other words, the two scenarios correspond to signals taking place in the early and later components, where the former is relatively easier for detecting difference than the later. In Section S.6 of Supplementary Material, we have included additional simulations that specifically focuses on signals occurring at large values of $K$ (e.g., $K=5,6$ for Scenario III and $K=7,8$ for Scenario IV). 
Besides, we also investigated the performance of proposed covariance test for different eigenvalue decay rates. 

To compare the empirical size and power, we conduct the three covariance tests in each scenario. For two groups we use two sample sizes $n=m=100$ and $n=m=200$, respectively, and let the sampling frequency $N=M$ vary in $\{6, 15, 30\}$ representing the paradigms from sparse to dense. For each replication, we use random equal-splitting and compute $T_{pool}$ twice by reversing the split samples. 
For the statistics $T_{FH}$ and $T_{PK}$ that are only applicable for fully observed functions, we run a local linear smoother 
through each subject and then use the pre-smoothed curves to compute them. The rejection rates are calculated for $1000$ and $500$ replications under the null and alternative hypotheses, respectively. The nominal significance level for all the tests is set at $0.05$.

\begin{figure}[!bpt]
	\centering
	\includegraphics[width=0.95\textwidth]{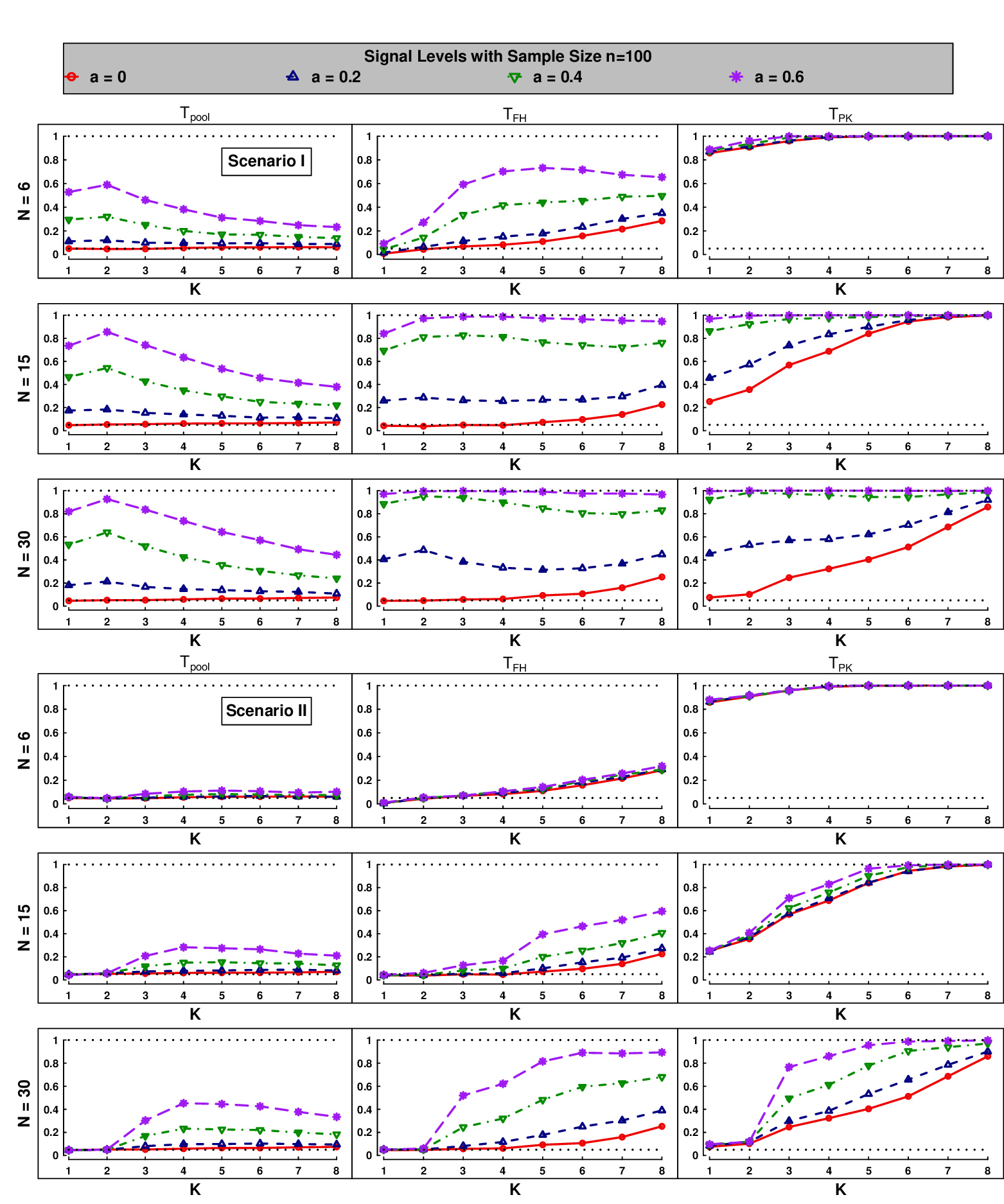}
	\caption{Empirical sizes and powers under Scenarios I and II (two sub-figures). Rejection rates of three statistics (from left to right) with sampling frequencies $N=M=6,15,30$ (from top to bottom) under $n=m=100$ are depicted at the significant level $0.05$ (dashed line). In each panel, power curves across different signal levels of $a=0, 0.2, 0.4, 0.6$ vary with truncation level $K$ from $1$ to $8$. In all cases, the tests are repeated $1000$ times for size and $500$ times for power.}
	\label{Fig1}
\end{figure}

The rejection rates of size $100$ are plotted in Figure \ref{Fig1}, while the figure of size $200$ are deferred to Figure S.3 of Supplementary Material for space economy. From the results, our simulations reveal distinct performance patterns across different testing settings. Overall, as $N$ and $n$ increase, all tests show the consistency across all scenarios. In the following, we discuss the simulated results in detail.

\textbf{(1) Empirical sizes ($a=0$):} From the empirical sizes (red line), we see that for small truncation levels (e.g., $K$ from $1$ to $3$), the sizes of proposed test $T_{pool}$ align well with the nominal significant level of $0.05$. Although a slight inflation occurs as $K$ increases, this trend reduces substantially with growing $N$ and $n$ (compared with Figure S.3 of Supplementary Material), which verifies our theoretical findings, i.e., detecting more components requires more measurements and subjects.
In contrast, the competitors $T_{FH}$ and $T_{PK}$ both exhibit obviously inflated sizes as $K$ increases (the later inflates more sharply). On small truncation levels (e.g., $K$ from $1$ to $3$) and under sparse designs (e.g., $N=6$), $T_{FH}$ has conservative size and $T_{PK}$ suffers from severe size inflation, however, their sizes tend to align with the significant level as $N$ and $n$ increase.

\textbf{(2) Powers analysis ($a>0$):} For the rejection rates representing the power, the proposed test $T_{pool}$ achieves the highest power when all anomaly signals are detected ($K=2$ in Scenario I and $K=4$ in Scenario II). However, power gradually deteriorates as $K$ increases further, which is expected since the test includes more unstably estimated high-order components for larger $K$. Across all scenarios, the competitor test $T_{FH}$ is nearly powerless compared to $T_{pool}$ under sparse diagram---consistent with the motivation of our approach. However, $T_{FH}$ slightly outperforms $T_{pool}$ for dense data, as it leverages the information of full asymptotic covariance matrix. In contrast, $T_{PK}$'s apparently higher power seems unreliable due to its severe size inflation, particularly in sparse designs. All tests show greater power in Scenario I than Scenario II, reflecting the inherent challenges of detecting smaller components (e.g., signals on larger $K$). Under weak signals (blue line with $a=0.2$), all tests exhibit lower power across most settings, and notably, $T_{FH}$ achieves highest power at $N=30$ in Scenario I due to its more efficient use of available information.


\section{Application to Longitudinal Diffusion Tensors} \label{realdata}

We apply the proposed covariance test procedure to the analysis of longitudinal diffusion tensors from Alzheimer's Disease Neuroimaging Initiative (ADNI, see \url{www.adni-info.org} for more information). After standard preprocessing procedures on the raw images, each diffusion tensor imaging (DTI) is derived in the form of a $3\times 3$ symmetric positive-definite matrix with eigenvalues $\lambda_1\geq\lambda_2\geq\lambda_3>0$. 
A simple and useful scalar measure of magnitude is the average of eigenvalues $\bar{\lambda}=(\lambda_1+\lambda_2+\lambda_3)/3$, referred to as the mean diffusivity (MD), which represents the total amount of diffusion in a voxel, i.e., related to the amount of water in the extracellular space \citep{OW2011}. We focus on the hippocampus, a brain region that is central to Alzheimer's disease, including $171$ subjects (at least four DTI scans) with age from $55.2$ to $93.5$. Among them, $n=39$ subjects are cognitively normal (denoted by CN) and the others $m=132$ developed one of (early or late) mild cognitive impairment and Alzheimer's disease, denoted by AD. On average, there are $5.56$ DTI scans for each subject, showing that the data are rather sparsely recorded. 


We are interested in testing whether there is difference between the second-order structures of MD in CN and AD groups. Three covariance tests $T_{pool}, T_{FH}, T_{PK}$ are carried out, and we conduct sample splitting for $T_{pool}$ and take turns reversing the roles of subsets to obtain three statistic values. The $p$-values of three statistics with various truncation $K$ are summarized in Table \ref{Tab1}. We see that the $p$-values of $T_{pool}$ rapidly decreases to $0.027$ and below, as $K$ increases exceeding $4$. This indicates that the difference of the covariance structures of MD trajectories in two groups may emerge in the subspace expanded by later principal components. By contrast, the $p$-values of test based on $T_{FH}$ are close to the nominal level $0.05$ at $K=3,4$ while the others are very large. The $p$-values of $T_{PK}$ are zeros across all $K$.

\begin{table}[h]
\renewcommand{\arraystretch}{0.8} 
\setlength{\tabcolsep}{9pt}
\caption{The $p$-values of statistics $T_{pool}, T_{FH}, T_{PK}$ applied to two groups CN and AD, with different truncation level $K$.}
\vskip 0.4cm
\centering
\scalebox{0.95}{
\begin{tabular}{ccllllllll}
\hline
\multicolumn{2}{l}{$K$}                                     & \multicolumn{1}{c}{1} & \multicolumn{1}{c}{2} & \multicolumn{1}{c}{3} & \multicolumn{1}{c}{4} & \multicolumn{1}{c}{5} & \multicolumn{1}{c}{6} & \multicolumn{1}{c}{7} & \multicolumn{1}{c}{8} \\ \hline
\multicolumn{1}{l}{$p$-value} & $T_{pool}$                  & 0.468                 & 0.488                 & 0.233                 & 0.244                 & \textbf{0.027}        & 0.006                 & 0.004                 & 0.004                 \\
                              & $T_{FH}$                    & 0.289                & 0.339                 & 0.056                 & 0.054                 & 0.212                & 0.474                 & 0.846                & 0.986                 \\
                              & $T_{PK}$                    & 0.000                 & 0.000                 & 0.000                 & 0.000                 & 0.000                 & 0.000                 & 0.000                 & 0.000                 \\ \hline
\end{tabular}
}
\label{Tab1}
\end{table}

\begin{figure}[!bpt]
	\centering
    \includegraphics[width=0.85\textwidth]{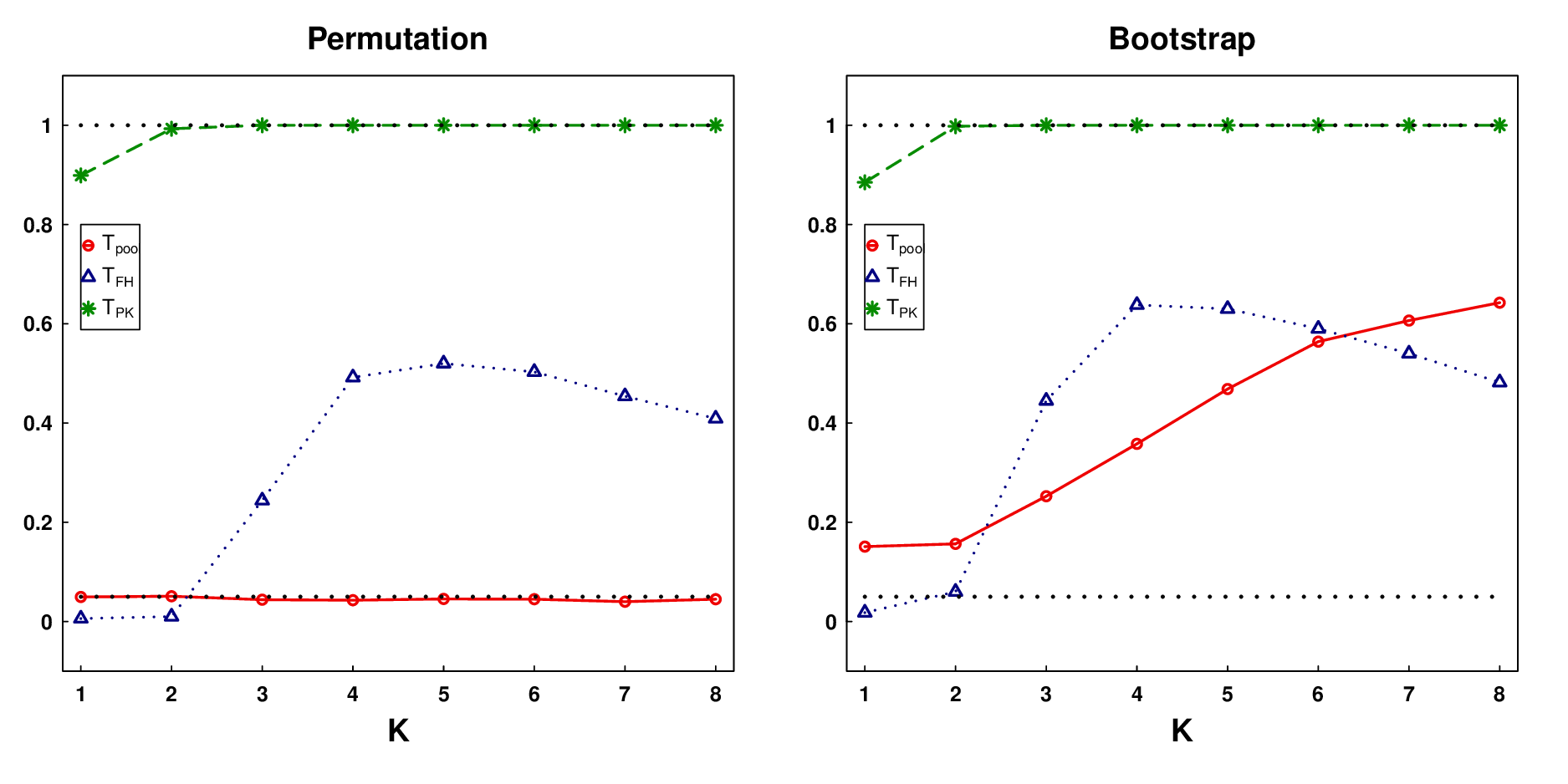}
    \caption{The rejection rates with increasing $K$ of three statistics $T_{pool}, T_{FH}, T_{PK}$ performed over $1000$ random permutations of the whole dataset (left) and $1000$ bootstrapped dataset (right), are depicted at the nominal significance level $0.05$ (dashed line).}
    \label{Fig-2}
\end{figure}

Although the assumptions in Section \ref{theory} are commonly imposed in FDA, it is difficult to verify them in practice. For justification, we employ resampling methods to exam the reliability of our method. Specifically, we perform $1000$ random permutations on the whole dataset by mixing up two groups, which eliminates the group difference and resemble the null hypothesis. Three tests are conducted over each permuted dataset and the results are shown in Figure \ref{Fig-2}. It is seen that the rejection rates of $T_{pool}$ agree well with the nominal level $0.05$, while those of the two competing methods are inflated substantially either over all $K$ (i.e., $T_{PK}$) or as $K$ goes up (i.e., $T_{FH}$). We also carry out $1000$ times random bootstrap with replacement separately within each group, generating two bootstrapped sets with sizes $n=39$ and $m=132$, to verify the power. The rejection rates for the three tests are shown in Figure \ref{Fig-2}. The rejection curve of $T_{pool}$ seems reasonable comparing to the $p$-values obtained from the dataset itself, while $T_{FH}$ seems to contradict with the $p$-values in Table \ref{Tab1}. The high rejection level of $T_{PK}$, together with the large size in the left panel under permutation, indicates to some extent that the significance suggested by $p$-values in Table \ref{Tab1} might be misleading. We demonstrate that the trajectory patterns of the three tests in each picture seem aligned with the performance in Scenario II under $N=6$ in Section \ref{simulation}. This further verifies that the difference occurs at smaller components for this dataset containing sparse observed functional data.

\section{Conclusion}\label{discussion}

In this paper, we delve into functional covariance test for discretely observed functional data under a \textit{nonparametric} framework, which remains unexplored since the existing works are mainly developed for fully observed functional data, including FPC-based tests 
with \textit{fixed} dimensions that undermines the nonparametric principle.
We utilize pool-smoothing covariance estimation to handle discretely observed data ranging from ``sparse'' to ``dense'' paradigms.
Additionally, by employing diagonal-elimination and sample-splitting strategy, the bias and variance are well controlled when the number of estimated eigenfunctions potentially to grow with sample size.
By investigating the asymptotic behaviour of a non-standardized statistic under the infinite-dimensional framework, we propose a test statistic by normalizing the variances of estimated scores in finite dimensions and derive its asymptotic Chi-squared null distribution with increasing truncations, saying $K=o\left(\min\{N^{1/(2\alpha+2)}, n^{1/(2\alpha+6)}\}\right)$.
This slowly increasing rate implies that in practice we can investigate the performance by running the test over a suitable range of different $K$ and makes a comprehensive and valid justification by the trend of testing performance. 
As in our simulations, for instance, we recommend to investigate $K$ up to $6$ under sparse diagram, and a wider range of $K$ up to $8$ for dense case. The extensive numerical experiments suggest that the proposed test is consistent as the sample size and the sample frequency increase.





\begin{center}
	{\large\bf Supplementary Materials}
\end{center}
The supplementary material provides all technical proofs and additional simulation results.


\begin{center}
{\large\bf Acknowledgments}
\end{center}
The authors would like to thank the editor, associate editor, and the anonymous referees for their insightful comments, which have significantly improved the article.

\begin{center}
{\large\bf Disclosure Statement}
\end{center}
The authors report there are no competing interests to declare.

\begin{center}
{\large\bf Funding}
\end{center}
Fang Yao's research is partially supported by the National Natural Science Foundation of China (No. 12292981), the National Key R\&D Program of China (No. 2022YFA1003800),
the New Cornerstone Science Foundation through Xplorer Prize, the LMAM, the Fundamental Research Funds for the Central Universities and the LMEQF.
Yang Zhou's research is partially supported by the National Natural Science Foundation of China (No. 12401334).



\bibliographystyle{asa}
\bibliography{TestCov}

\end{document}